\documentclass[
amsmath, %
floatfix, %
twocolumn, %
reprint, %
superscriptaddress, %
prb, %
aps, %
nofootinbib,%
nobibnotes,%
longbibliography, %
lengthcheck, %
final, %
]{revtex4-2} 
\renewcommand{\thispagestyle}[1]{}

\usepackage[T1]{fontenc}
\usepackage[utf8]{inputenc}
\usepackage{graphicx}
\usepackage{latexsym}
\usepackage{amsthm}
\usepackage[low-sup]{subdepth}

\usepackage{color}
\usepackage{mathtools}

\usepackage{placeins}

\usepackage[dvipsnames]{xcolor}

\usepackage[tighter]{newtxtext} 
\usepackage[subscriptcorrection,nosymbolsc,smallerops,bigdelims]{newtxmath} 
\DeclareMathAlphabet{\mathcal}{OMS}{cmsy}{m}{n} 
\DeclareMathAlphabet{\mathbcal}{OMS}{cmsy}{b}{n} 
\usepackage{bm}

\usepackage{floatrow}
\usepackage{siunitx}
\usepackage{hyperref}
\hypersetup{
	colorlinks,
	linkcolor={blue!90!black},
	citecolor={blue!90!black},
	urlcolor=	{blue!90!black}
}

\let\oldeqref\eqref
\renewcommand*{\eqref}[1]{%
	\hyperref[#1]{\oldeqref{#1}}%
}

\newcommand{\rr}{\bm{r}}

\newcommand{\kp}{{\bm{k} {\cdot} \bm{p}}}

\mathchardef\mhyphen="2D

\newcommand{\mr}[1]{\mathrm{#1}}

\newcommand{\upa}{{\uparrow}}
\newcommand{\doa}{{\downarrow}}

\DeclarePairedDelimiter\lr{\lparen}{\rparen}

\DeclarePairedDelimiter\abs{\lvert}{\rvert}

\DeclarePairedDelimiterX{\comm}[2]{\lbrack}{\rbrack}{#1, #2}

\DeclarePairedDelimiterX{\braket}[2]{\langle}{\rangle}{#1\delimsize\vert #2}
\DeclarePairedDelimiterX{\ketbra}[2]{\rvert}{\lvert}{#1 \delimsize\rangle\!\delimsize\langle #2}
\DeclarePairedDelimiterX{\matrixel}[3]{\langle}{\rangle}{#1 \delimsize\vert #2 \delimsize\vert #3}

\makeatletter
\newcommand{\raisemath}[1]{\mathpalette{\raisem@th{#1}}}
\newcommand{\raisem@th}[3]{\raisebox{#1}{$#2#3$}}
\makeatother

\definecolor{cbred}{HTML}{e31a1c}
\definecolor{cbgreen}{HTML}{33a02c}
\definecolor{cbblue}{HTML}{176aa7}

\usepackage{dcolumn}
\newcolumntype{d}[1]{D{.}{.}{#1}}

\newcommand{\figref}[1]{Fig.~\ref{fig:#1}}
\newcommand{\subfigref}[2]{Fig.~\hyperref[fig:#1]{\ref*{fig:#1}(#2)}}
\newcommand{\subfigsref}[3]{Figs.~\hyperref[fig:#1]{\ref*{fig:#1}(#2)}-\hyperref[fig:#1]{\ref*{fig:#1}(#3)}}

\newcommand{\bsubfigref}[2]{Figure~\hyperref[fig:#1]{\ref*{fig:#1}(#2)}}
\newcommand{\bsubfigsref}[3]{Figures~\hyperref[fig:#1]{\ref*{fig:#1}(#2)}-\hyperref[fig:#1]{\ref*{fig:#1}(#3)}}

\newcommand{\efield}{\mathcal{E}}
\newcommand{\efieldv}{\mathbcal{E}}

\newcommand{\bfield}{B}

\definecolor{cbred}{HTML}{e31a1c}
\definecolor{cbgreen}{HTML}{33a02c}
\definecolor{cbblue}{HTML}{176aa7}
\definecolor{cborange}{HTML}{ff7f00}
\definecolor{cbviolet}{HTML}{6a3d9a}

\newcommand{\upcapt}{\vspace{-1em}}

\usepackage[activate={true,nocompatibility} ,final ,tracking=alltext ,kerning=true ,spacing=true ,factor=1100 ,stretch=10 ,shrink=10 ,selected=true ,letterspace=0]{microtype}

\begin{document}
\title{Tunneling-related electron spin relaxation in self-assembled quantum-dot molecules}

\author{Micha{\l} Gawe{\l}czyk}
	\email{michal.gawelczyk@pwr.edu.pl}
	\altaffiliation{Present address: Institute of Physics, Faculty of Physics, Astronomy and Informatics, Nicolaus Copernicus University, Toru\'n 87-100, Poland}
	\affiliation{Department of Theoretical Physics, Faculty of Fundamental Problems of Technology, Wroc\l{}aw University of Science and Technology, 50-370 Wroc\l{}aw, Poland}
	\affiliation{Department of Experimental Physics, Faculty of Fundamental Problems of Technology, Wroc\l{}aw University of Science and Technology, 50-370 Wroc\l{}aw, Poland}

\author{Krzysztof Gawarecki}
	\affiliation{Department of Theoretical Physics, Faculty of Fundamental Problems of Technology, Wroc\l{}aw University of Science and Technology, 50-370 Wroc\l{}aw, Poland}

\begin{abstract}
We study theoretically spin relaxation during phonon-assisted tunneling of a single electron in self-assembled InAs/GaAs quantum-dot molecules formed by vertically stacked dots. We find that the spin-flip tunneling rate may be as high as $\SI{1}{\percent}$ of the spin-conserving one. By studying the dependence of spin relaxation rate on external fields, we show that the process is active at a considerable rate even without the magnetic field, and scales with the latter differently than the relaxation in a Zeeman doublet. Utilizing a multiband $\kp$ theory, we selectively investigate the impact of various spin-mixing terms in the electron energy and carrier-phonon interaction Hamiltonians. As a result, we identify the main contribution to come from the Dresselhaus spin-orbit interaction, which is responsible for the zero-field effect. At magnetic fields above $\SI{\sim15}{\tesla}$, this is surpassed by other contributions due to the structural shear strain. We also study the impact of the sample morphology and determine that the misalignment of the dots may enhance relaxation rate by over an order of magnitude. Finally, via virtual tunneling at nonzero temperature, the process in question also affects stationary electrons in tunnel-coupled structures and provides a Zeeman-doublet spin relaxation channel even without the magnetic field.
\end{abstract}

\maketitle

\section{Introduction}
The investigation of dynamics of spin degrees of freedom in semiconductor nanostructures, mainly quantum dots (QDs), is motivated both by still not fully explored and understood physics of spin relaxation and decoherence as well as the potential role of such structures in spintronics and quantum information processing \cite{RecherPRL2000,LossPRA1998}. Provided spin lifetimes of confined carriers are shown to be extraordinarily long, a promising application in spin-based memories additionally drives the interest \cite{KroutvarNature2004}. While various types of QDs can be utilized for these purposes, self-assembled structures come with the possibility of optical control as they are optically active, as opposed to, e.g., gate-defined structures. This allows one to perform essential operations of initialization, manipulation, and readout by means of light \cite{DuttPRL2005,GreilichPRL2006,AtatureScience2006,KronerPRL2008,XuNatPhys2008,RamsayPRL2008}, which provides fast operation. Recently, much attention is focused on tunnel-coupled structures like QD molecules \cite{XiePRL1995,KrennerPRL2005,StinaffScience2006}, as the coupling may be exploited in promising protocols developed for quantum information processing \cite{EconomouPRB2012,WeissPRL2012}.

While there is a considerable amount of theoretical works concerning spin relaxation of both types of carriers in single QDs of various types \cite{KhaetskiiPRB2001,WoodsPRB2002,WestfahlPRB2004,ChengPRB2004,ZipperJPhysCondensMatter2011,WeiPRB2012,LiSolidStateCommun2014}, the field of coupled QDs is far less explored in this context. Single existing studies refer to the impact of the presence of the second QD with the resultant tunnel coupling and other effects on the spin relaxation in one of the QDs for the electron \cite{StavrouJPhysCondensMatter2015} and the hole \cite{SegarraJPhysCondensMatter2015}. Regarding phonon-assisted tunneling in quantum-dot molecules, although the physics of the orbital transition itself has been widely recognized \cite{WuPRB2005,LopezRichardPRB2005,StavrouPRB2005,ClimentePRB2006,GrodeckaGradPRB2010,GawareckiPRB2010}, studies concerning the behavior of spin states during tunneling are scarce. They spotlight mostly on phonon-induced spin pure dephasing during tunneling \cite{GawelczykPRB2018,GaweczykAPPA2018} with a single perturbative estimate of the rate of spin-orbit-induced spin-flip tunneling \cite{GawelczykSST2017}.

Here, we focus on the phonon-assisted tunneling of electrons between coupled quantum dots and calculate the rate of such a transition affected by a simultaneous spin-flip as compared to the spin-conserving one. Employing a theoretical multiband $\kp$ framework, we calculate electron states as well as their coupling to the acoustic-phonon bath and the resultant rates of dissipative transitions. By studying the dependence of the spin-flip tunneling rate on external fields, we find that, even for an idealized structure, it may be as high as $\SI{1}{\percent}$ of the spin-conservative tunneling rate and its scaling with the magnetic field differs from that known from the Zeeman-doublet spin relaxation. Notably, the effect takes place also without the magnetic field at a considerable rate. By selectively enabling the spin-mixing terms in both electron energy and carrier-phonon interaction Hamiltonians, we quantify the impact of various mechanisms of spin relaxation. Those, via the effective conduction band description, can be associated with perturbative and phenomenological interaction terms known from the literature. We find that up to $B\mathbin{\sim}\SI{15}{\tesla}$ the main contribution to spin relaxation comes from the opposite-spin admixtures to the electron ground state caused by the Dresselhaus spin-orbit interaction. At the higher field, another source of admixtures becomes dominant, and that is the structural shear strain. Furthermore, we check the impact of typical details of the morphology of self-assembled structures on spin relaxation, and find that the misalignment of dots may elevate the zero- and low-field relaxation rate by order of magnitude.

Finally, we also find that at nonzero temperatures the process in question strongly enhances spin relaxation in Zeeman doublets in each of the QDs via virtual repetitive tunneling affected by spin-flips. In fact, we deal not only with a mere enhancement but with a channel of spin relaxation without the magnetic field that may set the limit for the spin lifetime of stationary carriers confined in tunnel-coupled structures utilized currently in spintronics and quantum information protocols \cite{EconomouPRB2012,WeissPRL2012}.

The paper is organized as follows. We begin in Sec.~\ref{sec:model} by specifying the system, describing the theoretical model, and providing general expectations regarding the results. Next, in Sec.~\ref{sec:sftunnel}, we present and analyze the calculated spin relaxation rates and their dependence on external fields as well as structural details of the quantum-dot molecule. Then, in Sec.~\ref{sec:thermal}, we show the crucial impact of studied transition on spin lifetime for stationary electrons in QDs forming the molecule. Finally, we conclude the paper in Sec.~\ref{sec:conclusions}. Additionally, in Appendices, we provide some more detailed information and results.

\section{System and model}\label{sec:model}
In this section, we characterize the system under study, describe the theoretical framework of our calculations, and construct some general predictions regarding the results.

\subsection{System}\label{subsec:system}
\begin{figure}[tb] %
	\begin{center} %
		\includegraphics[width=\columnwidth]{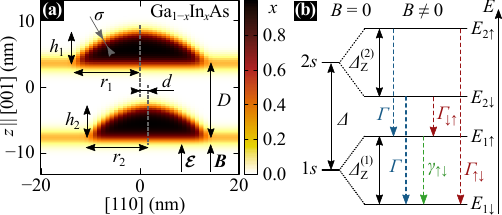} %
	\end{center} %
	\upcapt\caption{\label{fig:system}(a) A cross-section of the material composition (color gradient) of a double QD in the $(1\bar{1}0)$ plane. %
		(b) A schematic energy diagram of the system with relevant transitions marked with arrows.} %
\end{figure}
\begin{figure}[tb] %
	\begin{center} %
		\includegraphics[width=\columnwidth]{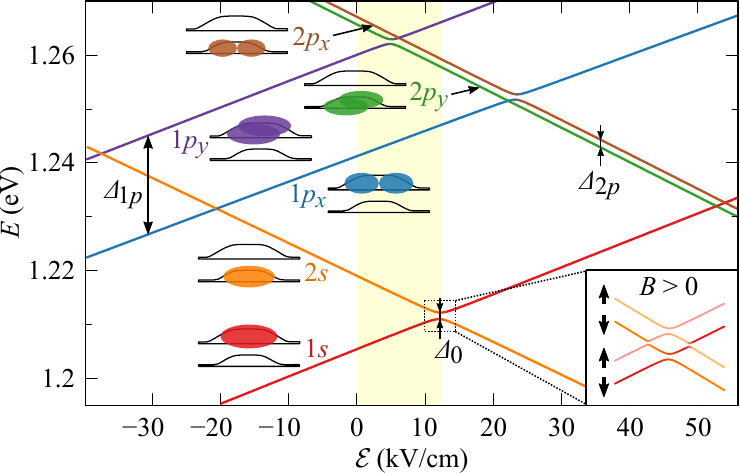} %
	\end{center} %
	\upcapt\caption{\label{fig:energy-E}Electric-field dependence of the lowest electron energy levels in the system at $B=\SI{0}{\tesla}$. Energy is given with respect to the unstrained GaAs valence-band edge. States are labeled by the dot number and orbital level, as shown in corresponding insets, and marked by line colors. Bottom right inset: schematic view of tunneling anticrossings at $\bfield > \SI{0}{\tesla}$.} %
\end{figure}
We model a self-assembled system of two vertically stacked flat-bottom, dome-shaped InAs QDs embedded in a GaAs matrix, separated by the distance $D=\SI{10.2}{\nano\metre}$. We assume a uniform
composition of \SI{100}{\percent} InAs inside the QD and the \SI{0.9}{\nano\metre}-thick wetting layer. The dots are of heights $h_1$ and $h_2$, respectively, and their base radii are $r_1$ and $r_2$. For most of the calculations, we assume equal height $h_1=h_2=\SI{4.2}{\nano\metre}$, and a slightly wider upper QD with $r_2=\SI{13.2}{\nano\metre}$ as compared to the bottom one with $r_1=\SI{12}{\nano\metre}$. We optionally account for the planar misalignment of QDs by a distance $d$, material diffusion at interfaces simulated by Gaussian averaging of the composition profile with the standard deviation $\sigma$, and unequal heights. These features represent typical morphological details met in layered self-assembled QDs \cite{KrennerNewJPhys2005}. A cross-section of the resulting material composition is presented in \subfigref{system}{a}, where QDs' dimensions are also given. Unless otherwise stated, we will refer to the basic model of sharp-interface coaxial QDs of equal height with $d=0$ and $\sigma=0$, and results are calculated at $T=\SI{0}{\kelvin}$. The structure is placed in axial electric $\efieldv=\efield\hat{z}$ and magnetic $\bm{B}=B\hat{z}$ fields.
In \figref{energy-E}, we present the calculated electric-field dependence of the lowest orbital electron levels. Apart from $s$-shell states, the transitions between which will be considered, $p$-type orbital levels are also included. Shaded area marks the range of electric-field magnitudes to be considered.

\subsection{Theoretical model}\label{subsec:model}
We follow a calculation scheme from structural modeling, through computation of the electron eigenstates, to the evaluation of the transition rates resulting from interactions with acoustic phonons. The strain distribution related to the InAs/GaAs lattice mismatch is calculated within the standard continuous elasticity approach \cite{PryorJAP1998}. Due to the lack of inversion symmetry in the zinc-blende crystal structure, the shear strain leads to the appearance of the piezoelectric potential. We account for the latter in a nonlinear regime by including the strain-induced polarization field up to the second-order terms in strain-tensor elements \cite{BesterPRL2006}, where the material parameters are taken from Ref.~[\onlinecite{CaroPRB2015}]. Having this included, we calculate the electron states using the eight-band envelope-function $\kp$ theory \cite{BurtJPhysCondMat1992,ForemanPRB1993}. While the inclusion of electric field is straightforward, we incorporate the magnetic field to the Hamiltonian by standard Zeeman terms and via the Peierls substitution in the gauge-invariant scheme \cite{AndlauerPRB2008}. Via numerical diagonalization, we obtain eigenvectors $\bm{\varPsi}\lr{\rr}$ in the form of pseudospinors, components of which are the envelope functions within each of the eight respective subbands. Those may be classified with the $\Gamma_{\mathrm{6c}}$, $\Gamma_{\mathrm{8v}}$, and $\Gamma_{\mathrm{7v}}$ irreducible representations of the ${T_{\mathrm{d}}}$ point group \cite{WinklerBOOK2003}. The $\Gamma_{\mathrm{6c}}$ block in the Hamiltonian is then related to the lowest conduction band, while $\Gamma_{\mathrm{8v}}$ to the heavy- and light-hole, and $\Gamma_{\mathrm{7v}}$ to the spin-orbit split-off subbands. The Dresselhaus spin-orbit interaction is accounted for via perturbative terms added to $H_{\mathrm{6c8v}}$ and $H_{\mathrm{6c7v}}$ Hamiltonian blocks responsible for coupling of conduction band to valence bands \cite{WinklerBOOK2003}. The $C_2$ parameter enters the off-diagonal Hamiltonian blocks with shear strain. For the calculations of the electron states, the position-dependent $C_2$ is introduced as in Ref.~[\onlinecite{Krzykowski2020}]. Finally, we include the impact of strain using the Bir-Pikus Hamiltonian with the standard $a_{\mathrm{c}}$, $a_{\mathrm{v}}$, $b_{\mathrm{v}}$, and $d_{\mathrm{v}}$ deformation potentials \cite{BirBOOK1974,BahderPRB1990}. The explicit form of the Hamiltonian, material parameters, and details of the numerical implementation of the model may be found in Ref.~[\onlinecite{GawareckiPRB2018}].

To investigate spin relaxation due to phonon-assisted dissipative transitions, we take into account the interaction of electrons with the acoustic-phonon bath in the long-wavelength limit via the deformation-potential (DP) and piezoelectric (PE) couplings \cite{GrodeckaBOOK2005}, both having the form of corrections to the Hamiltonian due to the phonon-induced strain field. The latter is plugged into the Bir-Pikus Hamiltonian to account for the DP coupling \cite{WoodsPRB2004,RoszakPRB2007}, while the PE interaction takes effect via the shear-strain-induced polarization field. Here, material parameters for GaAs are used.
Finally, the rates of transitions are found using the Fermi golden rule.

\subsection{Initial qualitative considerations}\label{subsec:qualitative}
The electron-phonon interaction is spin-conserving and hence there is no direct phonon-induced coupling of electron states with opposite spins. However, such spin-off-diagonal terms appear in the effective conduction-band description after including coupling to the valence bands via the $H_{\mathrm{6c8v}}$ and $H_{\mathrm{6c7v}}$ Hamiltonian blocks as well as spin-mixing terms in the valence-band blocks $H_{\mathrm{8v8v}}$ and $H_{\mathrm{8v7v}}$. Treating these couplings within the quasi-degenerate perturbation theory \cite{LowdinJChP1951}, one finds higher-order terms that induce electron spin relaxation \cite{MielnikPyszczorskiSR2018}. While they are caused by various effects, these spin-flip channels can generally be divided into two classes of \textit{admixture} and direct \textit{spin-phonon} mechanisms \cite{KhaetskiiPRB2000,KhaetskiiPRB2001}. 
Such distinction may be understood based on the separation into spin-conserving and perturbative spin-mixing parts, which is performed in both effective conduction-band Hamiltonians: the electron energy, $H_0+H_1$, and the carrier-phonon interaction, $V_0+V_1$. Then, up to the first order in perturbation, the overall effect is generated by pairs $H_1+V_0$ and $H_0+V_1$. The former is responsible for the admixture mechanisms: the perturbation $H_1$ prevents spin from being a good quantum number as each of electron eigenstates achieves an admixture from the opposite-spin subband. Such states can then be coupled by the fully spin-preserving interaction with phonons. The other combination describes the spin-phonon class of mechanisms, in which unperturbed electron states are coupled via the off-diagonal corrections to the interaction Hamiltonian (inherited after the Bir-Pikus $H_{\mathrm{6c8v}}$ and $H_{\mathrm{6c7v}}$ blocks). This may in fact be also thought of as dynamical induction of opposite-spin admixtures caused by phonons via the associated shear strain.

Calculation of spin relaxation rates within the Zeeman doublet in a single self-assembled QD may be found in Ref.~[\onlinecite{MielnikPyszczorskiPRB2018}] along with decomposition into various sub-mechanisms and evaluation of their contributions. There, the admixture caused by the structural shear strain proved to be the dominant source of spin-flips. Here, we deal with another transition involving spin relaxation, and our aim is to similarly assess which mechanisms are responsible for the effect. Understanding the sources of the two general spin relaxation channels allows us to have certain expectations. In the case of Zeeman doublet, one typically deals with the generic $\bfield^5$ magnetic-field dependence of relaxation rate. This is followed by all sub-mechanisms if PE coupling to phonons is involved. The only contributions exhibiting another, $\bfield^7$ scaling are those related to DP coupling within the admixture mechanism \cite{MielnikPyszczorskiPRB2018}. The difference of 2 in the powers of B may be traced back to the dependence of phonon spectral densities on transition energy: $E$ and $E^3$, respectively for PE and DP couplings, combined with the fact that for the doublet the whole energy difference is the Zeeman splitting, which proportional to $B$. The remaining power of 4 is related to the field-induced breaking of time-reversal symmetry. Let us consider a perturbation that generates the opposite-spin admixture to the electron eigenstate. In the presence of time-reversal symmetry, the two states have to be reflections of one another, so the admixtures are of exactly equal magnitude and opposite phases. Therefore,  their contributions cancel exactly under the matrix element of the carrier-phonon interaction Hamiltonian, and hence do not lead to a transition. However, in the presence of a magnetic field the symmetry is broken, which yields a difference in admixture magnitudes in the two states with the leading contribution proportional to $B^2$, which is reflected in the relaxation rate.

Here, we deal with tunneling characterized by finite transition energy, to which the Zeeman splitting forms only adds. Moreover, as the phonon spectral density for tunneling (or simply the tunneling rate), treated as a function of transition energy, is modulated by an oscillation with a frequency related to the interdot distance \cite{ClimentePRB2006,WijesundaraPRB2011} ($\mathbin{\propto}1/D$, see the inset in \figref{sf-rel-E}), it is highly ambiguous how such a change in the transition energy affects the rate. Since the energy separation is on a single meV scale, we may also expect a more pronounced role of the DP coupling, while the PE contribution should be of lesser importance, contrarily to the Zeeman-doublet case. Moreover, the two states connected by the transition are located in different dots, i.e., they are not the same orbital state, hence there is no need to break the time-reversal symmetry for the admixture mechanism to take place. Thus, one may expect a nonzero relaxation rate without the magnetic field, and a dependence on $B$ different from the one observed for the case of the Zeeman doublet. This will be assessed with the help of calculated $B$-field dependence of relaxation rates.

\section{Spin relaxation during tunneling}\label{sec:sftunnel}
In this section, we focus on the calculated rates of tunneling with spin relaxation by analyzing their dependence on external fields as well as determining the underlying mechanisms, and their relative contributions to the overall effect. 

\subsection{Spin-flip tunneling rates}\label{subsec:efield}
\begin{figure}[tb] %
	\begin{center} %
		\includegraphics[width=\columnwidth]{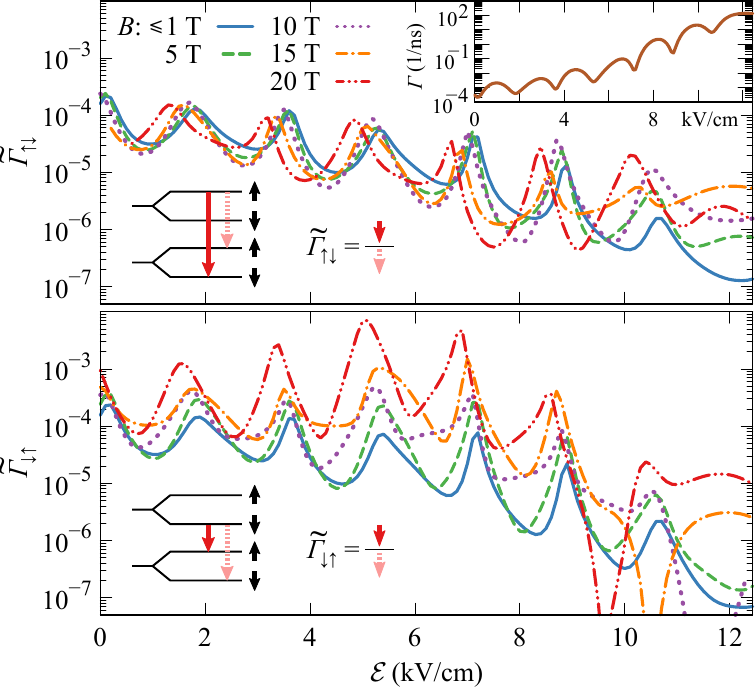} %
	\end{center} %
	\upcapt\caption{\label{fig:sf-rel-E}The ratio of the spin-flip-accompanied to spin-preserving electron tunneling rates, $\widetilde{\varGamma}_{\upa\doa}$ (top panel) and $\widetilde{\varGamma}_{\doa\upa}$ (bottom), as a function of the axial electric field $\efield$ at various magnitudes of the magnetic field $B$. Inset: Spin-preserving tunneling rate $\varGamma$ as a function of $\efield$. Note that $\varGamma\simeq\varGamma_{\mr{tot}}$, the total tunneling rate, as $\varGamma_{\upa\doa/\doa\upa}$ are orders of magnitude smaller.} %
\end{figure}
Having our initial predictions in mind, let us focus on the calculated rates of spin-flip tunneling. It is reasonable to consider, as a figure of merit with practical meaning, the relative rate, i.e., the ratio of rates for tunneling with ($\varGamma_{\upa\doa}$ or $\varGamma_{\doa\upa}$) and without ($\varGamma$) a spin-flip
\begin{eqnarray}\label{eq:rate-relative}
	\widetilde{\varGamma}_{\upa\doa/\doa\upa} = \frac{ \varGamma_{\upa\doa/\doa\upa} } { \varGamma },
\end{eqnarray}
rather than the bare rate of the former, as these two processes always compete with each other. One deals in practice with partial loss of spin polarization that is well quantified by this measure. Here, the subscript refers to the two opposite spin-flip processes [see \subfigref{system}{b} for definitions of all the rates].

To begin, we present in \figref{sf-rel-E} the dependence of the two relative spin relaxation rates (top and bottom panels) on the axial electric field at various magnitudes of the magnetic field. The first thing to notice is that we plot a common curve for all $B\mathbin{\le}\SI{1}{\tesla}$, as below this field magnitude they are almost constant. This means that we deal with phonon-assisted spin relaxation without the magnetic field, which is one of our main results and is discussed in detail later. Next, we note that while nearly identical at low magnetic field, the two rates evolve differently with increasing $\bfield$. In both cases, the uneven and shifted oscillations originate from the interplay of oscillations in the spin-preserving tunneling rate $\varGamma$ (extracted selectively from the calculation and plotted in the inset), which are virtually independent of $B$, with those in $\varGamma_{\upa\doa/\doa\upa}$. The latter are shifted as the transition energies change with rising Zeeman splitting $\varDelta_{\mr{Z}}^{(i)}=g_i\mu_{\mr{B}}B$,
\begin{align}
	\varDelta_{\upa\doa} = \varDelta + \frac{ \varDelta_{\mr{Z}}^{(1)} + \varDelta_{\mr{Z}}^{(2)} }{2}~, \quad 
	\varDelta_{\doa\upa} = \varDelta - \frac{ \varDelta_{\mr{Z}}^{(1)} + \varDelta_{\mr{Z}}^{(2)} }{2},
\end{align}
where $\varDelta_{\mr{Z}}^{(1)}$ and $\varDelta_{\mr{Z}}^{(2)}$ are Zeeman splittings in the two dots [see \subfigref{system}{b}]. This results in such uneven oscillations of their ratio. Zeeman splittings are slightly different due to small $\SI{\sim8}{\percent}$ mismatch of electron $g$-factors in QDs, $g_1=2.47$ and $g_2=2.29$, so transition energies for spin-preserving tunneling also shift marginally, $\varDelta_{\doa\doa/\upa\upa} = \varDelta \pm ( \varDelta_{\mr{Z}}^{(1)} - \varDelta_{\mr{Z}}^{(2)} )/2$.

\begin{figure}[tb] %
	\begin{center} %
		\includegraphics[width=\columnwidth]{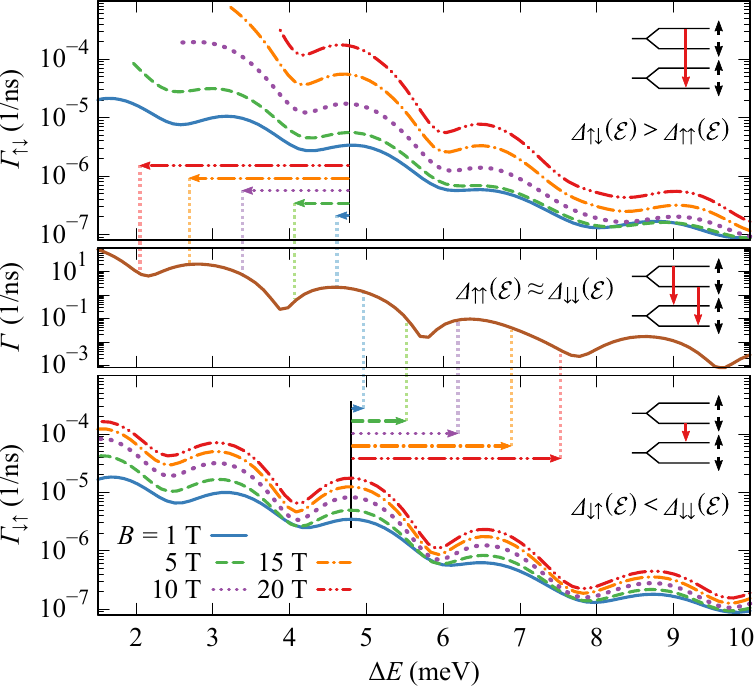} %
	\end{center} %
	\upcapt\caption{\label{fig:sf-abs-E}The rates of spin-flip (top and bottom) and spin-preserving (middle) electron tunneling, as a function of the energy difference $\Delta E$ for the given transition at various magnitudes of the magnetic field $B$ ($\varGamma$ is virtually $B$-independent). Horizontal arrows show how respective spin-flipping transitions are shifted with respect to spin-preserving ones in the magnetic field.} %
\end{figure}
With rising magnetic field a pronounced asymmetry between the relative rates of opposite spin-flips emerges, as $\widetilde{\varGamma}_{\upa\doa}$ is strongly enhanced but $\widetilde{\varGamma}_{\doa\upa}$ only shifts with a minor change in the value. This cannot be understood just based on the properties of electron-phonon interaction, as the one of the transitions that is relatively enhanced shifts with $B$ to higher transition energy, and the asymmetry also occurs in the range where phonon coupling is already decreasing. To inspect this, we show in \figref{sf-abs-E} the absolute rates, $\varGamma_{\doa\upa}$ and $\varGamma_{\upa\doa}$, of both spin-flipping transitions (top and bottom panels) along with $\varGamma$ (middle panel) plotted as a function of transition energy. Note that these curves are calculated for different values of $\efield$ and then brought to this common axis. Here, all oscillations are in phase, and both spin-flip rates increase with $\bfield$. Not only that, $\varGamma_{\upa\doa}$, which in relation to $\varGamma$, evolved much less in the magnetic field, here grows significantly faster than the other rate. This is however compensated for by energy shifts: as $\varDelta_{\doa\upa}\lr{\efield} < \varDelta_{\doa\doa}\lr{\efield}$ and $\varDelta_{\upa\doa}\lr{\efield} > \varDelta_{\upa\upa}\lr{\efield}$, the two transitions at a given electric field (i.e., in a real physical situation) coincide with $\varGamma$ values that may differ by orders of magnitude. This is indicated with arrows that mark the respective shifts between spin-flipping and spin-preserving transitions for each value of $B$, i.e., they show how the curves would be aligned on the $\efield$ axis, which has a physical meaning. Thus, we deal with two sources of the evolution of the relative rate in the magnetic field: relative shifts of transition energies proportional to the Zeeman splitting, and a more intrinsic dependence present also at fixed transition energy.

\begin{figure}[tb] %
	\begin{center} %
		\includegraphics[width=\columnwidth]{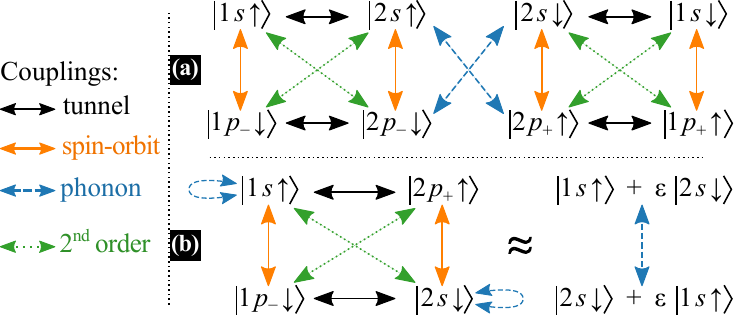} %
	\end{center} %
	\upcapt\caption{\label{fig:diagram}Schematic diagrams of selected couplings leading to phonon-assisted spin-flip tunneling transitions of admixture type in the system. In (b) a diagram for the opposite spin-flip process is obtained by flipping all spins and indices of $p$ states.} %
\end{figure}
The second effect, growth of $\varGamma_{\upa\doa }$ and $\varGamma_{\doa\upa}$ versus $B$ at a fixed $\Delta E$, is also uneven. To understand this, we need to focus on processes that allow for these spin-flipping transitions. As we show below in Sec.~\ref{subsec:bfield}, most of the spin-flip tunneling rate comes from admixture mechanisms announced above. To make the discussion easier to follow, in \figref{diagram} we sketch diagrams for some of the processes and couplings of that type, to which we attribute the observed transitions. Solid arrows depict couplings present in the electron Hamiltonian (tunnel and spin-orbit-like), dashed ones are for electron-phonon couplings, and diagonal dotted arrows are to show the effective second-order couplings originating from perturbative elimination of $p$-shell states. \bsubfigref{diagram}{a} shows the combination of regular spin-preserving tunnel coupling in both $s$ and $p$ shells, spin-orbit-induced opposite-spin $s$-$p$ shell mixing in each of QDs, and spin-preserving $p\mathbin{\to}s$ phonon-assisted relaxation \cite{GawelczykSST2017}. In \subfigref{diagram}{b}, we show another channel arising from spin-preserving $s$-$p$ tunnel coupling (enabled by broken axial symmetry) and spin-orbit interaction. In this case, the effective second-order coupling provides direct mixing of opposite-spin $s$-shell states in different QDs that may be coupled by phonons. The first diagram helps us explain the asymmetry observed in \figref{sf-abs-E}, as the mechanism involves the regular spin-preserving tunnel coupling. For a given energy difference, in the case of $\varGamma_{\upa\doa }$ we are closer to the tunneling resonance ($\upa\upa$), i.e., it occurs at higher $\efield$, than in the other case, and this difference is proportional to twice the average Zeeman splitting. If this is the reason, then an implicit inverse proportionality to this energetic distance from resonance, $\varDelta_{\upa\upa}-\varDelta_0$ for $\varGamma_{\upa\doa}$ and $\varDelta_{\doa\doa}-\varDelta_0$ for $\varGamma_{\doa\upa}$ ($\varDelta_0$ is the resonance width; see \figref{energy-E}), should be present in the rates. It is indeed the case, as multiplying both rates by such factors nearly cancels the difference.

Let us return to \figref{sf-rel-E} and note that at a low magnetic field both relative rates, apart from oscillations, increase when the distance to tunneling resonance increases (i.e., towards low $\efield$ values). While both spin-preserving and spin-flipping transitions are obviously getting weaker, their generally rising ratio may be explained by referring to the electric-field dependence of energy levels (see \figref{energy-E}), where the observed trend corresponds to approaching the weak $s\mhyphen p$ tunneling resonance. This may indicate a rising impact of processes of the type presented in \subfigref{diagram}{b}. Next, we notice that the most pronounced enhancement of $\varGamma_{\upa\doa}$ with $B$ takes place around $\efield=\SI{5}{\kilo\volt\per\centi\meter}$, which coincides with the tunneling resonance in the $p$-shell, which is, in turn, an essential part of processes depicted in \subfigref{diagram}{a}. Thus, these two tunnel couplings play an important role for spin relaxation, as they provide sources of spin-flipping state admixtures. One may notice that in the bottom panel, curves for $B = 10$, 15, and 20~T have dips at $\efield \simeq 11.5$, 10.5, and 9.5~kV/cm, respectively. They occur at the spin-flipping anticrossing that is shifted from the spin-preserving one by the Zeeman splitting. On both sides, we calculate tunneling in the energetically beneficial direction (which changes from QD1$\to$QD2 to QD2$\to$QD1 at this point).

There is one more asymmetry in the results, the rates at the low field are unequal on the two sides of the tunneling resonance, so for the opposite spatial directions of electron transfer. This difference gets reduced with rising $\bfield$, and we attribute it to the inequality of $p$-shell splittings in the QDs. We discuss it in Sec.~\ref{subsec:morphology}, where also a system of QDs of unequal height is studied. 

The dependence of investigated relative spin-flip tunneling rates on the electric and magnetic fields is complex and involves various factors. Using the electric field, the system may be driven close to important resonances that relatively enhance spin relaxation. The impact of the magnetic field may be coarsely divided into two main categories: the shift of the spin-flipping transition by the Zeeman splitting energy relative to the spin-preserving one and the more intrinsic effects involving the mixing of states. As the former effect is rather simple, it is reasonable to focus on the latter, which is not related to the specific character of the phonon spectral density for tunnel transitions. This, however, requires a more detailed analysis. Before we switch to it, let us emphasize the practical message of this section that $\widetilde{\varGamma}_{\doa\upa}$ can reach values as high as $\num{e-2}$.

\subsection{Magnetic-field dependence}\label{subsec:bfield}
\begin{figure}[tb] %
	\begin{center} %
		\includegraphics[width=\columnwidth]{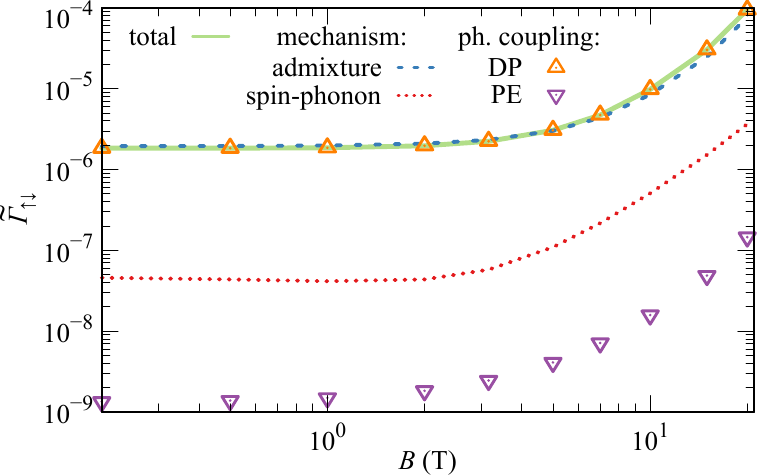} %
	\end{center} %
	\upcapt\caption{\label{fig:sf-rel-mech-B}The total relative spin-flip tunneling rate $\widetilde{\varGamma}_{\upa\doa}$ (solid line) compared to those arising from the admixture (dashed) and spin-phonon (dotted) mechanisms only, as well as divided into contributions due to the deformation-potential ({\large{\color{cborange}$\bm{\triangle}$}}) and piezoelectric ({\large{\color{cbviolet}$\bm{\triangledown}$}}) couplings to phonons, plotted as a function of the magnetic field $B$.} %
\end{figure}
To systematically examine the magnetic-field dependence of spin relaxation rates, we need a more synthetic figure of merit, which would be immune to variations due to moving along the oscillating spectral density with rising Zeeman splitting. The latter has an evident impact on $\widetilde{\varGamma}_{\upa\doa/\doa\upa}$ and is unavoidable in a real system, thus is of great practical importance. However, we would also like to reveal the more subtle impact via the influence of the magnetic field exerted by modifying the wave functions. Therefore, in \figref{sf-rel-mech-B}, we plot the ratio in which both of the rates, $\varGamma_{\upa\doa}$ and $\varGamma$ are calculated for the same transition energy. This is achieved by tuning the electric field to compensate for the Zeeman splitting. The choice of $\Delta E=\SI{4.812}{\milli\electronvolt}$ results from the fact that it corresponds to the last (closest to the tunneling resonance) maximum in $\varGamma$ that is available for spin-flipping tunneling at fields up to $B=\SI{20}{\tesla}$, as the spin-flipping resonance in this case shifts with $B$. Apart from the total relative rate (solid line), we also plot those arising from the admixture (dashed) and spin-phonon (dotted) channels treated separately. As predicted, there is a nonzero rate of spin relaxation at $B\mathbin{\to}\SI{0}{\tesla}$. Although it is not shown here, so as not to dim the picture, the value has been checked down to $B=\SI{e-4}{\tesla}$ and remains constant up to numerical precision. We find the total rate as dominated by the admixture channel, similarly to the case of spin relaxation in the Zeeman doublet \cite{MielnikPyszczorskiPRB2018}, with approximately $1.5$ orders of magnitude smaller spin-phonon contribution. It should be noted here that rates due to various mechanisms are not strictly additive, as interactions may also add up destructively, and the total effect may overcome the sum of individual contributions if those enter via nonlinear terms in the Hamiltonian. However, such a significant difference in magnitudes allows us to determine the dominant channel without a doubt.

Additionally, we split the total rate into the contributions due to DP and PE couplings to phonons (symbols). We deal with an almost negligible PE contribution to the rate generated by the DP coupling, which is understandable considering the transition energy exceeding the range of efficient PE interaction. The latter is subject to change for systems with narrower tunneling resonance, e.g., with larger interdot distance or bigger difference between the two dots. However, as this comes with strongly reduced tunneling rates, such structures are of lesser interest. Regarding the dependence on the magnetic field at its higher magnitudes, it is difficult to assess it for the total (and admixture-induced) rate. While it tends to scale as $B^4$ above $\SI{10}{\tesla}$, it is impossible to reproduce the moderate-field-range behavior with an $aB^4+c$ dependence. On the other hand, the spin-phonon-induced rate follows a well-defined $B^3+c$ trend. Thus, the unclear behavior has to come from the admixture contribution.

\subsection{Contributions to the relaxation rate}\label{subsec:contributions}
\begin{figure}[tb] %
	\begin{center} %
		\includegraphics[width=\columnwidth]{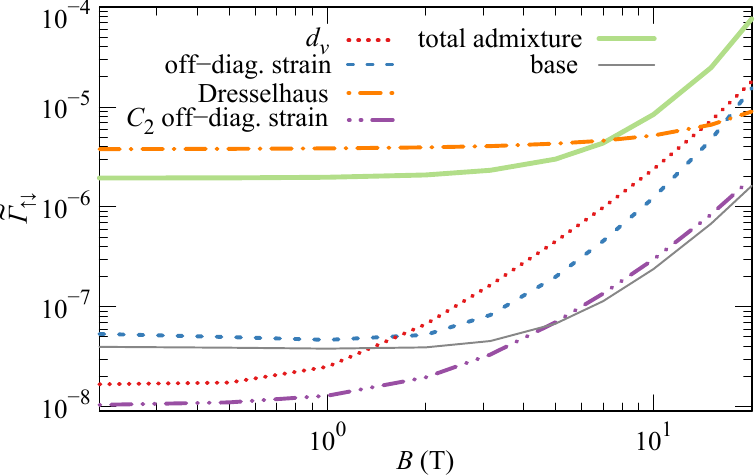} %
	\end{center} %
	\upcapt\caption{\label{fig:sf-rel-admix-B}The relative spin-flip tunneling rate due to the admixture mechanism (solid line) compared to the rates resulting from each of its constituent couplings within the valence band and in the band-off-diagonal blocks of the $\kp$ Hamiltonian, plotted as a function of the magnetic field $\bfield$.} %
\end{figure}
The opposite-spin admixtures, responsible here for most of the relaxation rate, may have multiple sources. For the case of relaxation within the Zeeman doublet, in large unstrained gate-defined QDs, the admixtures were found to originate mainly from the Dresselhaus spin-orbit interaction \cite{KhaetskiiPRB2001}. This is also the main reason for the spin-flip that accompanies electron $p\mathbin{\to}s$ orbital relaxation in such a kind of QDs \cite{KhaetskiiPRB2000}. On the other hand, in smaller self-assembled dots like those considered here, almost the entire rate was identified to originate from shear-strain-induced couplings between the conduction band and valence bands (via $H_{\mr{6c8v}}$ and $H_{6c7v}$ Hamiltonian blocks) as well as among the valence bands through terms proportional to the $d_v$ deformation potential in $H_{\mr{8v8v}}$ and $H_{\mr{8v7v}}$. In the effective description (model) for the conduction band, both these contributions turned out to have the form of a strain-induced correction to the electron Land{\'e} tensor, which is known from literature \cite{VanVleckPR1940,RothPR1960}. To find sources of spin-mixing admixtures in the case of tunneling transition, we plot in \figref{sf-rel-admix-B} the magnetic-field dependence of relative relaxation rates due to the admixture mechanism. Apart from the total rate, we also show contributions due to various couplings calculated by disabling all the respective matrix elements in the Hamiltonian except the one in question. One may notice that all the rates are non-vanishing at $B\mathbin{\to}0$ with the pronounced predominance of the Desselhus spin-orbit interaction in a wide range of fields up to $B=\SI{\sim 15}{\tesla}$. Via fitting with $aB^m+c$, we find an exponent of $m=2$, which, along with the order of magnitude of $\widetilde{\varGamma}_{\upa\doa}$, is in agreement with our recent perturbative calculation concerning this channel \cite{GawelczykSST2017}. At higher fields, the Dresselhaus contribution is surpassed by others, mainly those arising from shear-strain-induced couplings between the conduction and valence bands and among the latter (dashed and dotted lines, respectively). These are the same mechanisms that were found to dominate in the case of relaxation in the Zeeman doublet \cite{MielnikPyszczorskiPRB2018}. As both these strain-induced contributions have the effective form of corrections to the electron Land{\'e} tensor, their stronger scaling with the magnetic field is reasonable. Thus, the main difference compared to spin relaxation in a single QD is the strong zero-field contribution arising from the Dresselhaus spin-orbit interaction, which is absent in the relaxation within the Zeeman doublet. Among the $B$-dependent effects, i.e., after subtracting the zero-field values, strain-induced components are dominant. This is similar to the Zeeman doublet case. Here, it is even more pronounced than for a single QD due to the enhancement resulting from more delocalized electron wave functions (forming the bonding and antibonding states in the molecule), which thus penetrate more into the material interface, where shear strain is the highest. 

We need to stress out here that there are some residual spin-mixing terms that cannot be explicitly turned off in our calculation. These, containing mainly the Rashba spin-orbit coupling due to structural asymmetry, alone give rise to the rate plotted with a thin gray line labeled as ,,base''. Since this, each of the rates obtained for the other mechanisms carries such an implicit contribution. While it is negligible for the dominant Dresselhaus-induced rate, it may be at least partially responsible for the values of others at $B\mathbin{\to}0$. This would be in agreement with the interpretation of the shear-strain-induced contributions in terms of effective corrections to the $g$-factor, in view of which their impact should vanish at $B\mathbin{\to}0$.

The second class, spin-phonon mechanisms, was found to be responsible for the dominant Zeeman-doublet spin relaxation channel in large unstrained quantum dots \cite{KhaetskiiPRB2000}. Here it plays a minor role, similarly to the case of spin relaxation in single self-assembled dots \cite{MielnikPyszczorskiPRB2018}. Additionally, we find that the actual spin-phonon contribution may be even smaller than shown in \figref{sf-rel-mech-B}, as it is artificially enhanced by ,,base'' admixture contributions. We discuss this in more detail in Appendix \ref{app:spinphonon}.

\subsection{Impact of structure morphology}\label{subsec:morphology}
The results presented up to this point are obtained for the structure that may serve as a reference. It is reasonable to check how various details of structure morphology impact the spin-flip tunneling rates. We begin with evaluating the latter for varying in-plane size and height of QDs. Knowing that the Dresselhaus spin-orbit interaction is responsible for most of the effect and that its strength decreases with QD height (as $\langle k_z^2 \rangle \propto h^{-2}$), one could wish to alter it. However, taking the total rate as a figure of merit, we conclude that the impact of QD size is weak with the rate generally increasing with the volume of QDs, except for the low magnetic field case of height dependence, where a small drop of the total rate is present. The detailed results are presented in Appendix~\ref{app:morphology}. Thus, another strategy based on changing material composition of QDs may be used. For the studied material system, Dresselhaus spin-orbit interaction is weaker in the barrier GaAs material than in InAs, thus alloying the QDs with some amount of Ga could be used to reduce it. Going in that direction, alloyed InAs/InP and GaAs/AlGaAs systems are other options. On the other hand, e.g., the InSb/GaSb system comes with about an order of magnitude stronger spin-orbit interaction. However, one should note that in all these cases the nominal optical transition energy for a typical QD is different.

\begin{figure}[tb] %
	\begin{center} %
		\includegraphics[width=\columnwidth]{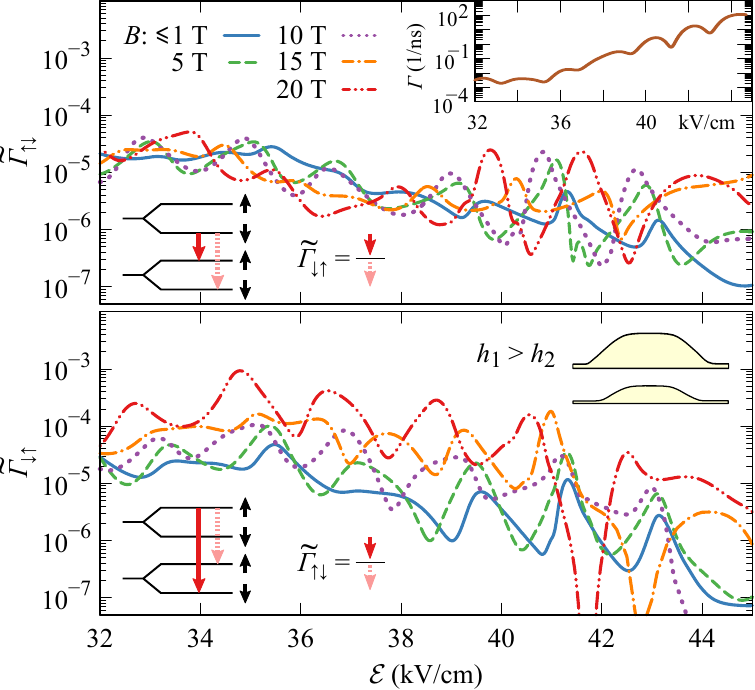} %
	\end{center} %
	\upcapt\caption{\label{fig:sf-rel-E-diffQDs}As \figref{sf-rel-E} but for the system with $h_1=\SI{4.2}{\nano\metre}$ and $h_2=\SI{2.4}{\nano\metre}$.} %
\end{figure}
Having analyzed the overall results obtained for the ideal structure, we now focus on the impact of typical morphological features met in coupled QDs. As the growth of the upper QD takes place in the strain field of the bottom one, its size may be significantly different \cite{BrackerAPL2006}. Such a system is nominally far from the $s$-shell tunneling resonance and an axial electric field with a magnitude of the order of tens of keV/cm has to be used to approach it. To see how this affects the studied transition rates, we show in \figref{sf-rel-E-diffQDs} an analog of \figref{sf-rel-E}, but calculated for a system with QDs of significantly different heights: $h_1=\SI{4.2}{\nano\metre}$ and $h_2=\SI{2.4}{\nano\metre}$. We set the bounds of applied $\efield$ to probe a similar range of energy splittings as previously.

Starting the analysis from the top inset, we notice that unequal heights partially destroyed the oscillation of spin-preserving tunneling rate $\varGamma$, especially at higher splittings (lower $\efield$). This translates into a similar effect in the low-$\efield$ part of results in both panels. However, at higher $\bfield$ the oscillations are recovered, which is better visible for $\widetilde{\varGamma}_{\doa\upa}$. Inspection of the absolute rates shows that oscillations for spin-flipping transitions are in fact intact. This reveals some nuances of spin-mixing admixtures. Decay of oscillations in $\varGamma$ comes from interfering terms arising from the different spatial extent of electron wave functions in the two QDs, and thus different spatial frequencies they decompose into in the reciprocal space. Oscillation damping is absent for the rates $\varGamma_{\upa\doa/\doa\upa}$, which are caused by admixtures, so we conjecture that the localization length of the latter has to depend on QD height very weakly. If we carefully analyze oscillations in $\varGamma$ and $\varGamma_{\doa\upa}$ for the reference system in \figref{sf-abs-E}, we notice a small \SI{\sim 3}{\percent} mismatch of their periods. This translates into a \SI{0.3}{\nano\meter} shorter interdot distance for the coupling via admixtures. This may be understood given that spin-orbit interaction enters via terms cubic in momentum. Thus, the resultant admixtures to the $s$-shell wave function need to be odd, and as the momentum operator is represented by real-space differentiation, their envelopes should be located where the main part of wave function drops, so closer to the interface, instead of the QD center. Inspection of oscillations for the unequal-height QD system confirms this conjecture, as we find a roughly twice as big mismatch of periods in this case ($\SI{\sim8}{\percent} \sim \SI{0.8}{\nano\metre}$). 

\begin{figure}[tb] %
	\begin{center} %
		\includegraphics[width=\columnwidth]{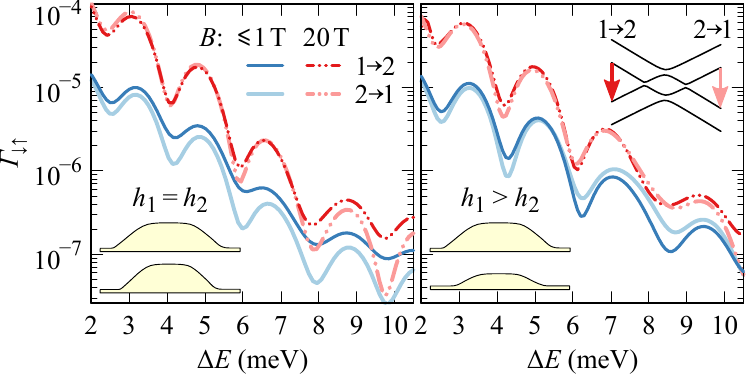} %
	\end{center} %
	\upcapt\caption{\label{fig:sf-abs-E-compare}The absolute rate $\varGamma_{\doa\upa}$ of spin-flip tunneling at the low and high magnetic field calculated for the two sides of the tunneling resonance (darker and lighter curves) for QDs of equal (left) and different (right) heights.} %
\end{figure}
As announced at the end of Sec.~\ref{subsec:efield}, the rates of spin-flip tunneling are not symmetric on the two sides of the resonance. It is useful to discuss this when comparing systems with equal and different dot heights. In \figref{sf-abs-E-compare}, we show the absolute rate $\varGamma_{\doa\upa}$ calculated on both sides: darker curves are for $\efield$ below, and lighter above the resonance, as a function of transition energy for the two systems (left and right panels). In both cases, the asymmetry present at the low magnetic field vanishes at $\bfield=\SI{20}{\tesla}$. Thus, it is the zero-field contribution that causes this imbalance. Additionally, it is stronger for the reference system than for the one with unequal QD heights. Seeking an explanation, let us focus on the big disproportion of $p$-shell splittings in the former (see \figref{energy-E}):  $\varDelta_{1p}\simeq\SI{19}{\milli\electronvolt}$ and $\varDelta_{2p}\simeq\SI{1.5}{\milli\electronvolt}$. This comes mostly from the strain-induced piezoelectric field, which typically affects the upper QD much more. Thus, the role of the piezoelectric field is important here. As a result, we deal with a large asymmetry of the energy diagram of the reference system: both the $p$-shell and $s$-$p$ tunneling resonances on the low-$\efield$ side are closer to the $s$-shell one than it is on the other side. Moreover, one of the admixture mechanisms depends directly on the $p$-shell splitting \cite{GawelczykSST2017}. On the other hand, in the unequal-height system splittings are much closer: $\varDelta_{1p}=\SI{11.7}{\milli\electronvolt}$ and $\varDelta_{2p}=\SI{8.1}{\milli\electronvolt}$ (see Appendix~\ref{app:morphology}, \figref{energy-E-diffQDs} for energy levels of this system). While the energy diagram is still not fully symmetric, all distances and splittings are comparable.

While the differences and nuances discussed above allowed us to get some deeper insight into the processes that underlay spin-flip tunneling transitions, the practical conclusion is that these two morphologically different systems behave similarly regarding studied transitions.

\begin{figure}[tb] %
	\begin{center} %
		\includegraphics[width=\columnwidth]{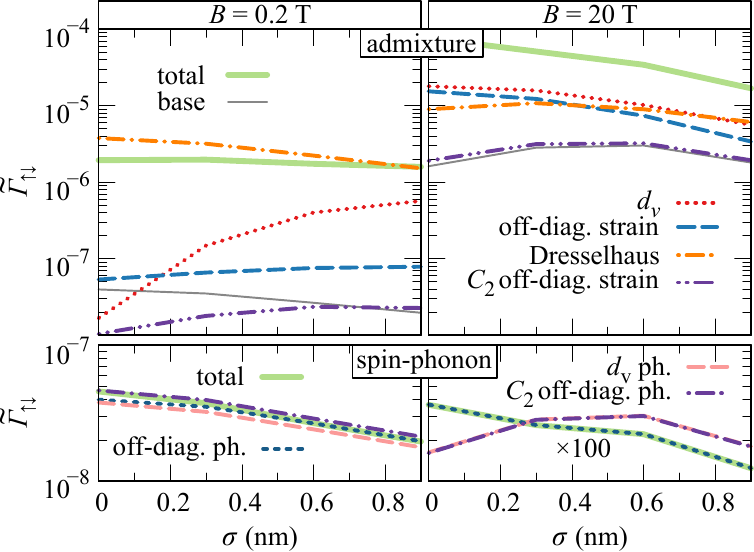} %
	\end{center} %
	\upcapt\caption{\label{fig:sf-rel-all-sigma}The relative spin-flip tunneling rates due to individual mechanisms and their combinations at $B=\SI{0.2}{\tesla}$ (left panels) and $B=\SI{20}{\tesla}$ (right panels) plotted as a function of the interfacial material intermixing length $\sigma$. The top (bottom) row of panels presents rates due to mechanisms from the admixture (spin-phonon) class.} %
\end{figure}
Another morphological detail of real systems is that the material interface of self-assembled QDs is never sharp due to the diffusion of atoms. We simulate this by Gaussian averaging of the material composition that leads to a soft interface with the characteristic length of intermixing $\sigma$. To assess how it affects the individual spin-relaxation mechanisms, we plot in \figref{sf-rel-all-sigma} the calculated relative rates versus $\sigma$ for two values of the magnetic-field magnitude $B=\SI{0.2}{\tesla}$ (left panels) and $B=\SI{20}{\tesla}$ (right) representing the low- and high-field regimes. We begin with analyzing the admixture mechanism. Although one could expect that softer interfaces should lead to a reduction of structural shear strain, at the low field we observe an increase of the rate related to the latter. While the expectation is basically correct, the most significant result of interface softening is the enhancement of penetration of the wave function into the barrier, where it experiences the impact of strain located at the interface. This is, however, compensated for by a decrease of the contribution due to the Dresselhaus spin-orbit interaction, and the total rate is approximately $\sigma$-independent. At the higher field, both shear-strain-induced rates are decreasing with $\sigma$, which may be understood as the barrier penetration is prevented in this case due to the in-plane shrinkage of wave functions in the magnetic field combined with reduced strain magnitude. Regarding the rates induced by spin-phonon mechanisms, we deal with a very weak impact of the interface softening.

\begin{figure}[tb] %
	\begin{center} %
		\includegraphics[width=\columnwidth]{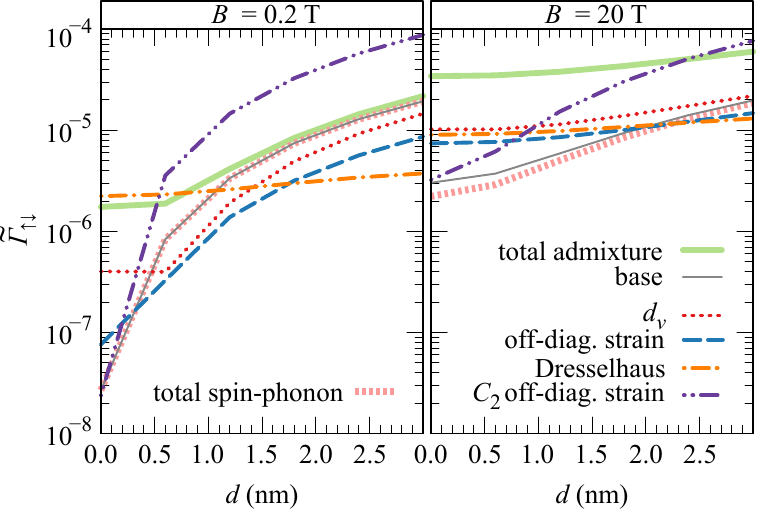} %
	\end{center} %
	\upcapt\caption{\label{fig:sf-rel-all-d}The relative spin-flip tunneling rates due to individual mechanisms and their combinations at $B=\SI{0.2}{\tesla}$ (left panel) and $B=\SI{20}{\tesla}$ (right panel) plotted as a function of the planar misalignment of the dots $d$.} %
\end{figure}
Another feature of self-assembled QD molecules, unavoidably present in real samples, is the misalignment of the dots. In \figref{sf-rel-all-d}, we analyze the impact of the latter, quantified with the distance $d$ on the relative spin relaxation rates. While the effect at the high field is minor, we deal with a significant increase of all the admixture contributions in the low-field regime. This results from the lowering of the symmetry, which strongly enhances the $s\mhyphen p$ orbital tunneling resonances \cite{DanielsPRB2013,GawareckiPRB2014} that are crucial for one of the mechanisms of creation of spin-mixing admixtures for electrons [see diagram in \subfigref{diagram}{b}]. For comparison, we note here that in the case of holes, apart from this mechanism, breaking the axial symmetry also leads to opposite-spin admixtures through coupling to the light-hole subbands, which has been studied as the dominant mechanism \cite{DotyPRB2010,RajadellAPL2013}. In \figref{sf-rel-all-d}, we do not split the spin-phonon induced effects into sub-mechanisms as the difference between those rates is not noticeable in the scale of the overall variation of their values. However, we plot the total spin-phonon-induced rate to notice that at low $B$-field it is in fact caused by the residual admixtures (``base''), as mentioned above, as the two coincide perfectly.

\subsection{Relaxation at nonzero temperature}
Up to now, we considered the $T=\SI{0}{\kelvin}$ limit. Trivially, all rates of phonon-assisted transitions depend on temperature via the Bose distribution of phonon-mode occupations, , $n_{\mr{B}}\lr{\omega}$, which enters the expression for transition rate via the factor $\abs{n_{\mr{B}}\lr{\Delta E/\hbar,T}+1 }$. This applies to both spin-preserving and spin-flip tunneling, thus the ratio of their rates undergoes a weak dependence induced only by the mismatch of transition energies, which is equal the Zeeman splitting $\varDelta_{\mr{Z}}$,
\begin{equation}\label{eq:rate-fixed-dE}
	\left. \widetilde{\varGamma}_{\upa\doa/\doa\upa}\right\rvert_T = \left. \widetilde{\varGamma}_{\upa\doa/\doa\upa}\right\rvert_{T=\SI{0}{\kelvin}} \frac{\abs*{n_{\mr{B}}\lr*{\frac{\varDelta\pm\varDelta_{\mr{Z}}}{\hbar},T}+1}}{\abs*{n_{\mr{B}}\lr{\varDelta/\hbar,T}+1}}.
\end{equation}
In the above, the sign distinguishes between opposite spin-flip processes, as $\widetilde{\varGamma}_{\upa\doa}$ and $\widetilde{\varGamma}_{\doa\upa}$ are shifted towards higher and lower transition energies, respectively. The factor on the right-hand side saturates with temperature, at low and moderate magnetic fields to a value close to 1. At $\bfield=\SI{20}{\tesla}$ the room-temperature values are about $0.67$ and $2.25$, respectively, for the given DQD with $g\mathbin{\approx}2.4$ and $\varDelta$ set to $\SI{4.812}{\milli\electronvolt}$. Thus, with rising temperature, the relative overbalance of the $\doa\upa$ relaxation becomes enhanced. 

\section{Enhancement of Zeeman-doublet spin relaxation}\label{sec:thermal}
In this section, we show that the process of spin-flip tunneling investigated here affects electron spin not only during the transition but has also an impact on the stationary electron located in one of the QDs, which leads to an additional channel of spin relaxation in the Zeeman doublet.

\begin{figure}[tb] %
	\begin{center} %
		\includegraphics[width=\columnwidth]{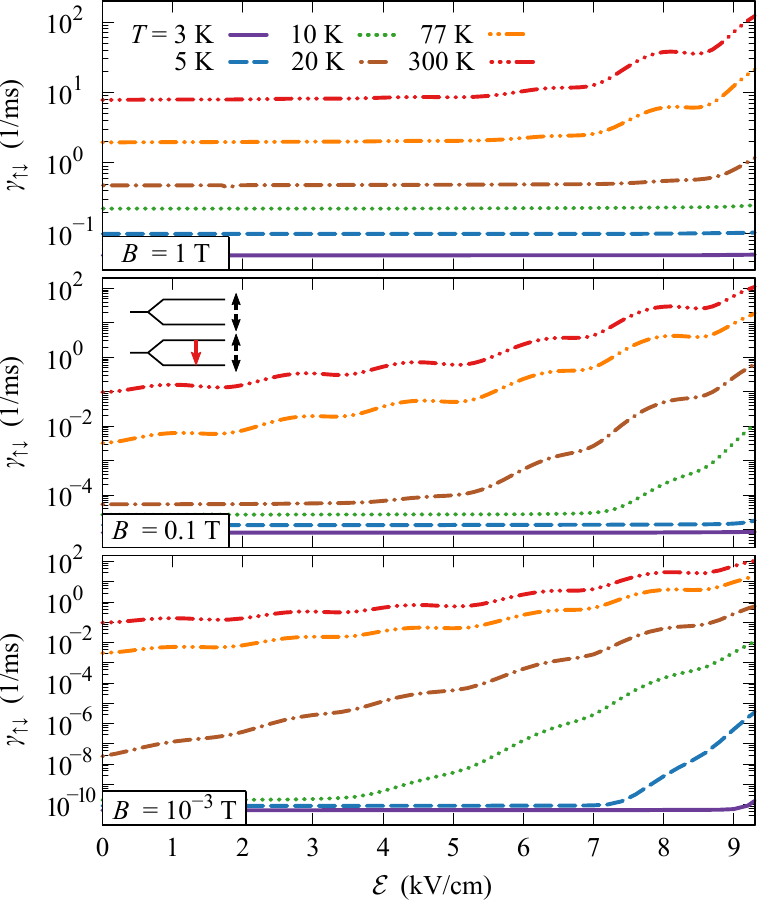} %
	\end{center} %
	\upcapt\caption{\label{fig:sf-abs-cumulative-E}The rate of spin relaxation in the orbital ground state Zeeman doublet, $\gamma_{\upa\doa}=1/T_1$ , in the lower-energy QD, plotted as a function of the axial electric field $\efield$ at various temperatures $T$, for $B=\SI{1}{\tesla}$ (top panel), $B=\SI{0.1}{\tesla}$ (middle) and $B=\SI{e-3}{\tesla}$ (bottom).} %
\end{figure}
Let us consider an electron that does not actually tunnel, but stays approximately in its orbital ground state in the lower-energy QD. Exposed to the phonon-assisted tunnel coupling to the other dot, such an electron occupies at a finite temperature a mixture of states localized in the two dots, which is dominated by the ground state, according to the detailed balance condition. Provided the tunneling rates in both directions are finite, at equilibrium the electron continuously undergoes a virtual process of repetitive tunneling between the dots, which is affected by spin-flips of the nature discussed in previous sections. This additional channel should enhance spin relaxation in the ground-state Zeeman doublet. To quantify the effect, we consider a set of rate equations for the four-level system, to which we insert the numerically calculated rates of all spin-preserving and spin-flipping transitions. Solving these via the Laplace transform method, we find an exponential component in the ground-state spin evolution, which describes the effective spin relaxation within the Zeeman doublet. In \figref{sf-abs-cumulative-E}, we plot the resulting spin relaxation rates $\gamma_{\upa\doa}=1/T_1$ for three values of magnetic field: $\bfield=\SI{e-4}{\tesla}$, \SI{0.1}{\tesla}, and \SI{1}{\tesla} at application-relevant temperatures as a function of the axial electric field in the previously considered range close to the tunneling resonance. Starting from the highest magnetic-field case, we notice plateaus where the rate is weakly enhanced compared to the bare direct spin-flip (equal to the low-$\efield$ plateau values). At elevated temperatures, the rate increases as the system get pushed towards the tunneling resonance with the rising electric field. The lower the magnetic field, the more pronounced the role of the discussed spin relaxation channel is, which is simply due to the vanishing rate of the direct spin-flip at $B=0$. Importantly, close to the resonance, the calculated rate very weakly depends on $B$ and provides a spin relaxation channel at $B\mathbin{\to}0$ with rates reaching $\SI{\sim e-2}{\milli\second^{-1}}$ at $T=\SI{5}{\kelvin}$ and over $\SI{\sim e2}{\milli\second^{-1}}$ at $T=\SI{300}{\kelvin}$.

\begin{figure}[tb] %
	\begin{center} %
		\includegraphics[width=\columnwidth]{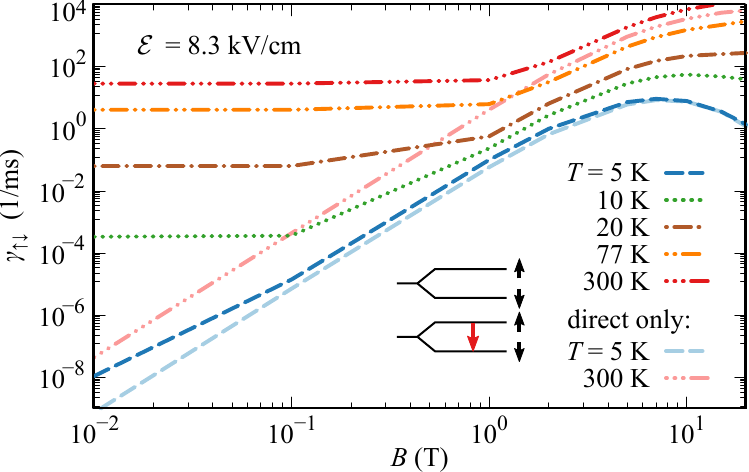} %
	\end{center} %
	\upcapt\caption{\label{fig:sf-abs-cumulative-B}The rate of spin relaxation in the ground-state Zeeman doublet, $\gamma_{\upa\doa}=1/T_1$, in the lower-energy QD, plotted as a function of the magnetic field $B$ at various temperatures $T$, for $\efield=\SI{8.3}{\kilo\volt\per\centi\meter}$. Lighter lines show rates of the bare direct phonon-induced spin relaxation for selected temperatures.} %
\end{figure}
For a better insight, we set the electric field to $\efield=\SI{8.3}{\kilo\volt\per\centi\meter}$, corresponding to the local maximum of the tunneling rate [see the inset to \figref{sf-rel-E}], and in \figref{sf-abs-cumulative-B} we plot the effective spin relaxation rates as a function of the magnetic field at various temperatures. Confronting the effective rates with those of direct spin-flip (plotted with gray lines for the two most outlying of simulated temperatures) we may notice how the virtual-tunneling channel plays a dominant role in the low and moderate field range. This is on top of the standard $B^5$ direct spin-flip rate that has a maximum at $B\sim\SI{8}{\tesla}$ and then drops (as the transition energy equal to Zeeman splitting crosses the maximum of phonon spectral density due to piezoelectric coupling). Hence, we deal with a phonon-induced Zeeman-doublet spin relaxation without the magnetic field, as opposed to all direct channels that exhibit power laws in the $B$ dependence and hence vanish at $B=0$.

\section{Conclusions}\label{sec:conclusions}
We have theoretically investigated the electron confined in a self-assembled double QD system and presented the calculated rates of tunneling transition with a simultaneous spin-flip as compared to the spin-preserving process. Using a multiband $\kp$ theory and including the coupling to acoustic phonons with all leading-order spin-perturbing effects included, we have calculated the electron states and the rates of transitions between them. By checking the dependence of the investigated spin-flip tunneling rate on external fields, we have determined that it can reach \SI{1}{\percent} of the spin-preserving one for an idealized structure, which may be further increased with structural asymmetry. Most importantly, the rate does not vanish even at $B\mathbin{\to}{0}$. Our theoretical framework allowed us to selectively turn on various spin-mixing terms both in electron energy and interaction Hamiltonians, so we could assess the relative contributions of individual spin relaxation mechanisms. Unlike the Zeeman-doublet case studied before, we have found that the Dresselhaus spin-orbit interaction is responsible for most of the spin relaxation in a wide range of magnetic field magnitudes. At about $B=\SI{\sim15}{\tesla}$, it gets surpassed by the interactions induced by the structural shear strain, which, via the effective conduction-band description, may be understood as corrections to the electron $g$-factor. Considering the morphology of realistic self-assembled quantum-dot molecules, we have learned that planar misalignment of the dots strongly enhances the low-field spin relaxation rate by about an order of magnitude for a typical sample geometry. Finally, we have shown that at nonzero temperatures the studied process also leads to the Zeeman-doublet spin relaxation for stationary electrons via virtual tunneling to the other dot. This provides a phonon-related source of spin-flips at zero magnetic field, which crucially depends on temperature and may limit spin lifetime for carriers confined in tunnel-coupled structures.

\begin{acknowledgments}
We acknowledge support from the Polish National Science Centre under Grants Nos 2014/14/M/ST3/00821 (M.G.) and 2014/13/B/ST3/04603 (K.G.).
Calculations have been carried out using resources provided by Wroclaw Centre for Networking and Supercomputing \cite{wcss}, Grant No.~203.
We are grateful to Pawe{\l} Machnikowski for valuable discussions and advice.
\end{acknowledgments}

\FloatBarrier
\appendix
\section{Spin-phonon contributions}\label{app:spinphonon}
\begin{figure}[tb] %
	\begin{center} %
		\includegraphics[width=\columnwidth]{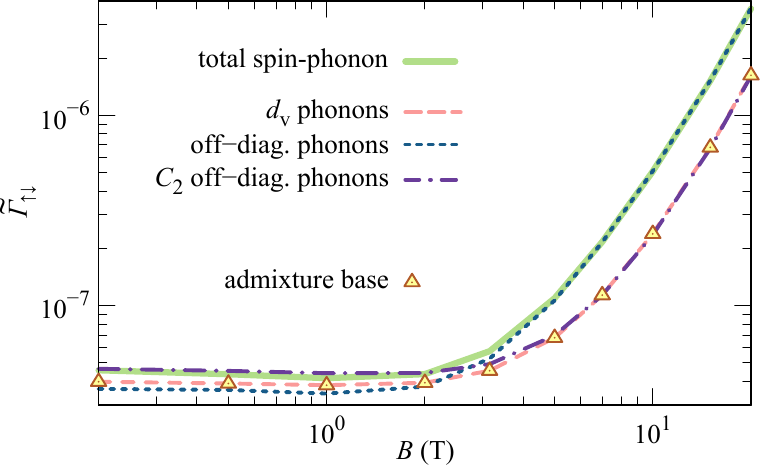} %
	\end{center} %
	\upcapt\caption{\label{fig:sf-rel-spinphonon-B}The relative spin-flip tunneling rate due to the spin-phonon mechanism (solid line) compared to the rates resulting from each of its constituent couplings, plotted as a function of the magnetic field $\bfield$. Additionally, points show the base admixture contribution.} %
\end{figure}
Here, for completeness, we analyze in detail the contribution of spin-phonon mechanisms, which was skipped in the main text due to minor relevance. In \figref{sf-rel-spinphonon-B}, we plot the relative spin relaxation rates due to various spin-phonon mechanisms. These split similarly to the strain-induced admixtures, as they enter via equivalent Hamiltonian matrix elements, but with a strain field coming from a different source: phonons instead of structural lattice deformation. From about $B=\SI{3}{\tesla}$ the rate is dominated by the contribution, in which the phonon-induced shear strain enters via the off-diagonal block of the interaction Hamiltonian that couples conduction band to valence bands. Similarly to the admixture case, this mechanism may be interpreted as a correction to the electron $g$-factor, which here is dynamically induced by the phonon shear-strain field. Regarding the low-field regime, it is ambiguous whether we really deal with a nonzero contribution at $B\mathbin{\to}0$. Let us recall that there are some sources of admixtures (mainly the Rashba spin-orbit coupling) that may not be switched off in our calculation. Those form the ,,base'' contribution to the admixture-induced spin relaxation rate, which is unavoidably present also here (although it is not a spin-phonon mechanism), and its contribution alone is plotted with symbols for comparison. In view of this, it is reasonable to assume that at least most of the low-field rate attributed to spin-phonon mechanisms is, in fact, due to the residual admixtures.

\section{Details of the impact of structure morphology}\label{app:morphology}
\begin{figure}[tb] %
	\begin{center} %
		\includegraphics[width=\columnwidth]{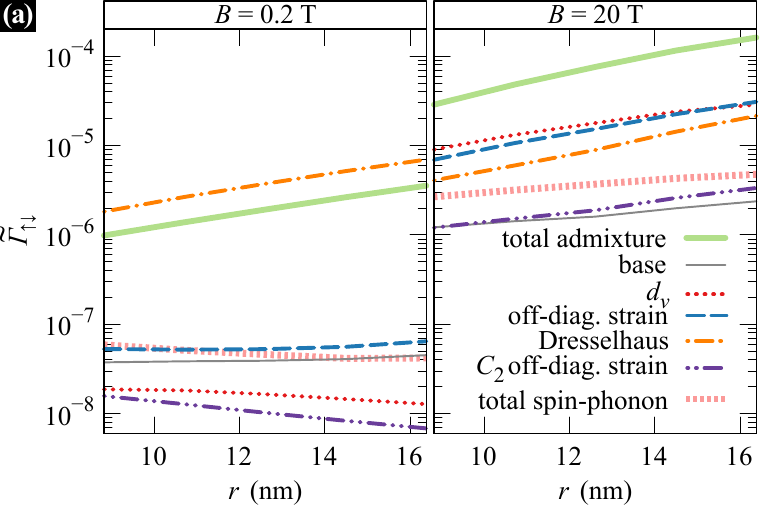}\\ %
		\vspace{4pt}
		\includegraphics[width=\columnwidth]{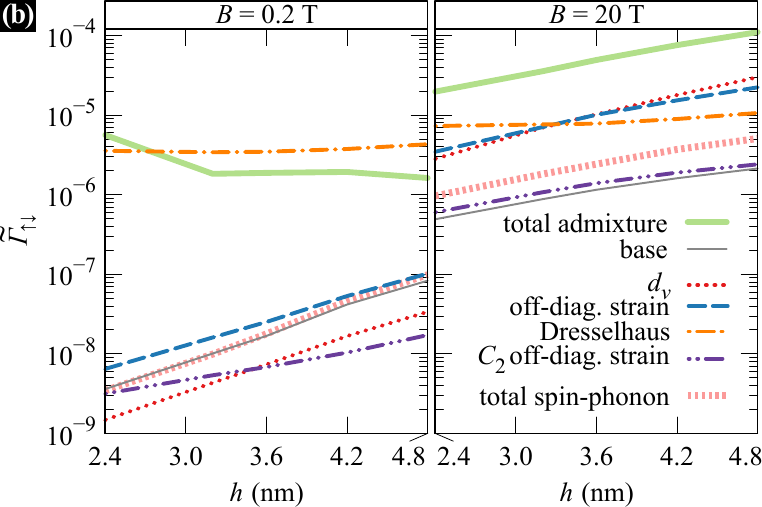} %
	\end{center} %
	\upcapt\caption{\label{fig:sf-rel-all-radius-height}The relative spin-flip tunneling rates due to individual mechanisms and their combinations at $B=\SI{0.2}{\tesla}$ (left panel) and $B=\SI{20}{\tesla}$ (right panel) plotted as a function of: (a) the average in-plane radius $r$ of the dots (while keeping the \SI{10}{\percent} mismatch, $r_2=1.1\,r_1$), and (b) the height $h = h_1 = h_2$ of the dots.} %
\end{figure}
As discussed in Sec.~\ref{subsec:morphology}, one could in principle expect the spin-flip tunneling rates to depend on the size of QDs, and the strength of this dependence is not easy to estimate based on qualitative considerations. Thus, we model series of structures, in which the radius and height of QDs are varied while keeping the dots similar. The results are presented in \subfigref{sf-rel-all-radius-height}{a} and \subfigref{sf-rel-all-radius-height}{b}, respectively. In general, taking the total rate as a figure of merit, the impact of varying sizes of QDs is weak. In both cases at the high magnetic field, we observe some enhancement of all contributions with increasing QD size. We attribute this to the reduction of level splittings in QDs, which in turn results in the creation of larger admixtures where coupling between $s$ and $p$ shells takes place. The only drop of the total relative spin-flip tunneling rate is present in the low-$B$ case of varying height. This appears to come from the cancellation of some contributions, as none of them presents such a drop itself in magnitude.

\begin{figure}[tb] %
	\begin{center} %
		\includegraphics[width=\columnwidth]{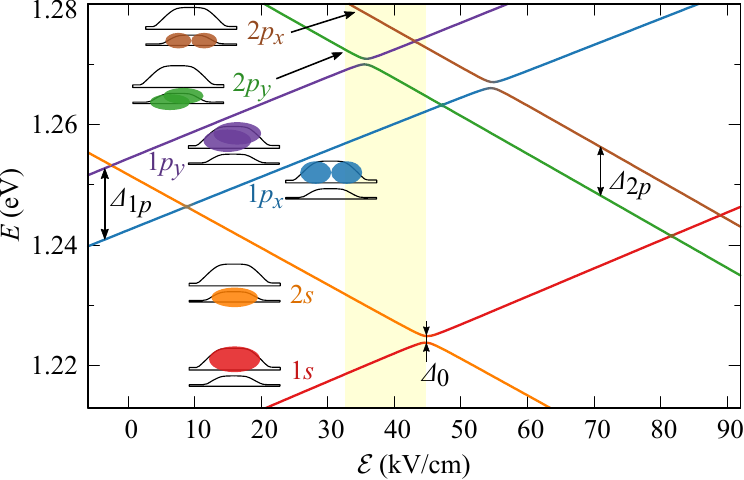} %
	\end{center} %
	\upcapt\caption{\label{fig:energy-E-diffQDs}As \figref{energy-E} but for QDs with $h_1\!=\SI{4.2}{\nano\metre}$ and $h_2\!=\SI{2.4}{\nano\metre}$.} %
\end{figure}

Finally, we show in \figref{energy-E-diffQDs} the electric-field dependence of energy levels for the system with unequal QD heights discussed in Sec.~\ref{subsec:morphology}. One may notice that the diagram is much more symmetric, and $p$-shell splittings are not as different as for the reference structure.

\FloatBarrier

%

\end{document}